\documentclass{article}
\usepackage{spconf,amsmath,graphicx}

\usepackage{enumitem}
\usepackage{bm}
\usepackage{lipsum}
\usepackage{amssymb}
\usepackage{amsmath}
\usepackage{xcolor}
\setlist{nosep, leftmargin=14pt}
\usepackage{soul}

\usepackage{mwe} 

\title{Semi-supervised Cervical Segmentation on Ultrasound by A Dual Framework for Neural Networks}
\name{Fangyijie Wang$^{\star \dagger}$ \qquad Kathleen M. Curran$^{\star \dagger}$ \qquad Gu\'enol\'e Silvestre$^{\star \S}$}

\address{
    $^{\star}$ Taighde \'{E}ireann – Research Ireland Centre for Research Training in Machine Learning  \\
    $^{\S}$ School of Computer Science, University College Dublin, Dublin, Ireland \\
    $^{\dagger}$ School of Medicine, University College Dublin, Dublin, Ireland
    }
\begin{document}
%
\maketitle

\begin{abstract}
Accurate segmentation of ultrasound (US) images of the cervical muscles is crucial for precision healthcare. The demand for automatic computer-assisted methods is high. However, the scarcity of labeled data hinders the development of these methods. Advanced semi-supervised learning approaches have displayed promise in overcoming this challenge by utilizing labeled and unlabeled data. This study introduces a novel semi-supervised learning (SSL) framework that integrates dual neural networks. This SSL framework utilizes both networks to generate pseudo-labels and cross-supervise each other at the pixel level. Additionally, a self-supervised contrastive learning strategy is introduced, which employs a pair of deep representations to enhance feature learning capabilities, particularly on unlabeled data. Our framework demonstrates competitive performance in cervical segmentation tasks. Our codes are publicly available on https://github.com/13204942/SSL\_Cervical\_Segmentation.
\end{abstract}


\section{Introduction}
Transvaginal ultrasound is the preferred method for visualizing the cervix in most patients as it provides detailed insight into cervical anatomy and structure \cite{Alcazar:2014}. Accurate segmentation of ultrasound (US) images of the cervical muscles is crucial for analyzing the deep muscle structures, evaluating their function, and monitoring customized treatment protocols for individual patients \cite{Bones:2024}.

The manual annotation of cervical structures in transvaginal ultrasound images is a labor-intensive and time-consuming process, which restricts the availability of extensive labeled datasets essential for building robust machine learning models. To address this challenge, semi-supervised learning (SSL) techniques have shown potential by incorporating labeled and unlabeled data, thus enhancing the extraction of valuable insights from unannotated data \cite{Engelen:2020,Yang:2023}. 

This study presents an SSL framework for cervical segmentation on ultrasound images. First, the SSL framework is designed for network training on many unlabeled data. Then, pixel-level cross-supervised learning is introduced within the framework. Therefore, the network is trained with the help of the other network via pseudo-labeling. A contrastive learning strategy is introduced within this framework, incorporating a pair of embedded features to maximize the feature learning capabilities using both labeled and unlabeled data. Moreover, the framework is validated using a dataset in a public challenge, demonstrating competitive performance.

\section{Methodology}
\begin{figure*}[htb]
  \centering
  \includegraphics[width=\linewidth]{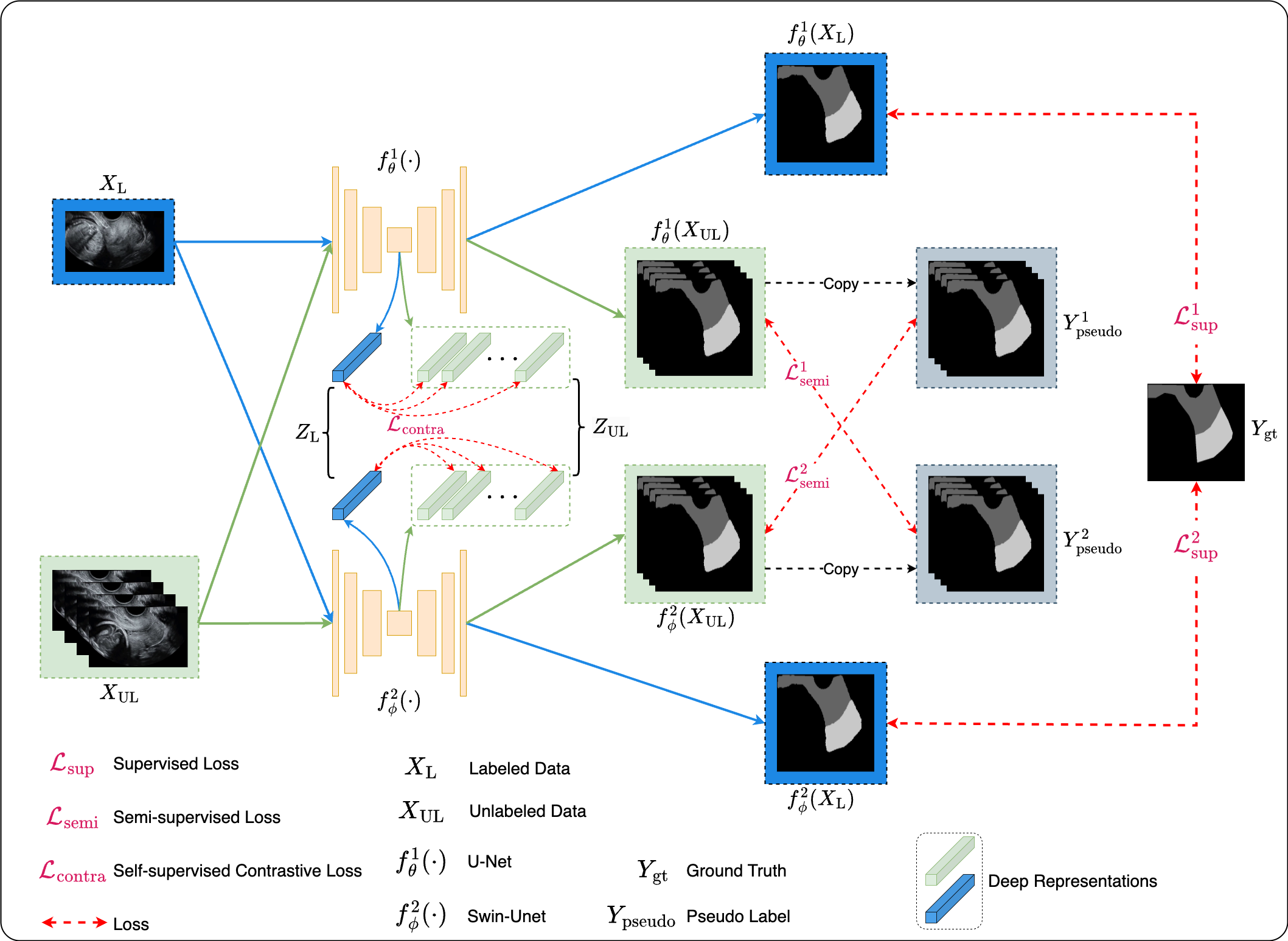}
  \caption{Framework for deep representations contrastive cross-supervised neural networks for semi-supervised ultrasound image segmentation.}  \label{ssl_framework}
\end{figure*}


The SSL framework of dual neural networks is illustrated in Fig.\ref{ssl_framework}, where $(\bm X_\text{L}, \bm Y_\text{gt}) \in \bm L$ denotes the labeled training dataset, while $(\bm X_\text{UL}) \in \bm U$ denotes the unlabeled training dataset. $\bm X_\text{L}, \bm X_\text{UL} \in \mathbb{R}^{3 \times n_h \times n_w}$ represents a 2D ultrasound image of size $n_h\times n_w$ with 3 channels.
The dual neural networks are denoted by $f^1_{\theta}(\cdot)$ and $f^2_{\phi}(\cdot)$, respectively. $\bm Y_\text{pseudo}, \bm Y_\text{gt} \in \mathbb{R}^{3 \times n_h \times n_w}$ represents a three-class labeled segmentation masks with pixel values ranging from 0 to 2. The segmentation masks predicted by the segmentation networks $f^1_{\theta}(\cdot)$ and $f^2_{\phi}(\cdot)$ are $f^1_{\theta}(\bm X)$ and $f^2_{\phi}(\bm X)$, respectively, with $\theta$ and $\phi$ as their parameters. 
In our proposed method, the prediction of a network is considered as a pseudo label to expand the unlabeled dataset to $(\bm X_\text{U}, \bm Y_\text{pseudo})$ to train the other network. We utilize each network to extract the deep representation features of $\bm X_\text{L}$ and $\bm X_\text{UL}$ for contrastive learning. To select and save the best network from $f^1_{\theta}(\cdot)$ and $f^2_{\phi}(\cdot)$, we evaluate their segmentation performance by measuring the difference between $(f^1_{\theta}(\bm X), \bm Y_\text{gt})$ and $(f^2_{\phi}(\bm X), \bm Y_\text{gt})$ in our validation set. 

Our training objective is minimizing the total loss $\mathcal L_\text{total}$ by updating the network parameters $\theta$ and $\phi$. The total loss $\mathcal L_\text{total}$ consist of the supervision loss $\mathcal L_\text{sup}$, semi-supervised loss $\mathcal L_\text{semi}$, and self-supervised contrastive loss $\mathcal L_\text{contra}$. Mathematically, it can be expressed as:
\begin{equation}
\label{loss}
\mathcal{L}_{\text {total }}=(\mathcal{L}_{\text {sup }}^1+\mathcal{L}_{\text {sup }}^2)+\lambda(\mathcal{L}_{\text {semi }}^1+\mathcal{L}_{\text {semi }}^2)+\mathcal{L}_{\text {contra }}
\end{equation}
where $\lambda$ represents the weighting factor for a ramp-up function used exclusively for the unlabeled training set \cite{Laine:2017}. This function facilitates the gradual transition of $f^1_{\theta}(\cdot)$ and $f^2_{\phi}(\cdot)$ from being initialized with the labeled training set to prioritizing learning from the unlabeled training set.
In Fig.\ref{ssl_framework}, all loss functions are highlighted by a red dashed line.   $\mathcal{L}^1_\text{sup}$ and $\mathcal{L}^2_\text{sup}$ are the supervision losses for $f^1_{\theta}(\cdot)$ and $f^2_{\phi}(\cdot)$ based on the labeled training set. $\mathcal{L}_\text{sup}$ is designed with a combination of the Dice Similarity Coefficient (DSC) and cross-entropy (CE) losses, as follows:

\begin{equation}
\begin{split}
\mathcal{L}_{\text {sup }}^1 &=\operatorname{CE}\left(f^1_{\theta}\left(\bm{X}_\text{L}\right), \bm{Y}_{\mathrm{gt}}\right) +\operatorname{DSC}\left(f^1_{\theta}\left(\bm{X}_\text{L}\right), \bm{Y}_{\mathrm{gt}}\right) \\
\mathcal{L}_{\text {sup }}^2&=\operatorname{CE}\left(f^2_{\phi}\left(\bm{X}_\text{L}\right), \bm{Y}_{\mathrm{gt}}\right) +\operatorname{DSC}\left(f^2_{\phi}\left(\bm{X}_\text{L}\right), \bm{Y}_{\mathrm{gt}}\right)
\end{split}
\end{equation}

\subsection{CNN and Transformer Segmentation Network}
The UNet \cite{Ronneberger:2015} presents a groundbreaking modification of the traditional encoder-decoder segmentation network, incorporating diverse network blocks tailored for medical image analysis. We adopt UNet as the convolutional neural network (CNN) $f^1_{\theta}(\cdot)$ within our framework. In recent years, several variants of UNet have demonstrated superior segmentation performance in medical imaging analysis \cite{AnbuDevi:2022}. We leverage a Swin-UNet architecture \cite{Cao:2023} as $f^2_{\phi}(\cdot)$ to implement our SSL strategy. The detailed architecture of Swin-Unet is elucidated in the study by Cao et al. \cite{Cao:2023}. We initialize both networks with random weights.

\subsection{Cross-Supervised Learning}
Inspired by consistency regularization and multi-view learning principles, including cross-pseudo-supervision \cite{chen:2021,Ma:2024}, this study applies these concepts to construct dual architectures that facilitate mutual learning. 
CNNs excel in learning spatial hierarchies of features, while Transformer-based networks excel in capturing broad, non-local interactions. Hence, we propose a simple yet powerful cross-supervised learning approach that combines CNNs and Transformers to assist each other mutually. The semi-supervised loss function, denoted as $\mathcal L_\text{semi}$, can be expressed as:


\begin{equation}
\begin{split}
\mathcal{L}_{\text {semi }}^1=\operatorname{CE}\left(f^1_{\theta}\left(\boldsymbol{X}_\text{UL}\right), \bm Y^2_\text{pseudo}\right) + 
\operatorname{DSC}\left(f^1_{\theta}\left(\boldsymbol{X}_\text{UL}\right), \bm Y^2_\text{pseudo}\right) \\
\mathcal{L}_{\text {semi }}^2=\operatorname{CE}\left(f^2_{\phi}\left(\boldsymbol{X}_\text{UL}\right), \bm Y^1_\text{pseudo}\right) +
\operatorname{DSC}\left(f^2_{\phi}\left(\boldsymbol{X}_\text{UL}\right), \bm Y^1_\text{pseudo}\right)
\end{split}
\end{equation}

We also investigated the approach of cross-teaching between two networks, including identical two CNNs or Transformer networks. Nevertheless, our proposed method, which integrates CNN and Transformer-based models, surpasses the effectiveness of this approach.

\subsection{Contrastive Learning}
Contrastive learning is widely acknowledged as a powerful paradigm for extracting robust and discriminative features, marking a significant advancement in the field of self-supervised learning \cite{Aaron:2019}. The fundamental concept of contrastive learning is that both positive and negative samples are discriminative. The utilization of contrastive learning in the field of medical image analysis enhances the capability of feature extraction, ultimately leading to improved model performance \cite{You:2022,Wang:2023,Ma:2024}. 

In the paper~\cite{He:2020}, contrastive learning can be considered as a task to search a dictionary. Given an encoded query $q$, a set of encoded keys $\{ k_1,k_2, \dots \}$ is retrieved from the memory bank. Among these keys, a specific positive key $k^+$ aligns with the query $q$, while the remaining negative keys $k^{-}$ represent different images.
A contrastive loss function InfoNCE \cite{Aaron:2019} is utilized to bring $q$ closer to positive key $k^+$ and simultaneously distance it from the negative keys $k^{-}$:
\begin{equation}
\mathcal L_q^\text{NCE}=-\log \frac{\exp \left(q \cdot k^{+} / \tau\right)}{\exp \left(q \cdot k^{+} / \tau\right)+\sum_{k^{-}} \exp \left(q \cdot k^{-} / \tau\right)}
\end{equation}
where $\tau$ denotes a temperature hyper-parameter. In our SSL framework, we refer to the original InfoNCE loss, which is formalized as follows:

\begin{equation}
\mathcal{L}_{\text {contra }}=\sum_{i}^{\text{L}} \sum_{j}^{\text{UL}} \text{InfoNCE}(Z^i_\text{L}, Z^j_\text{UL})
\end{equation}
where $\text{InfoNCE}$ represents the original InfoNCE loss function, $Z^i_\text{L}$ and $Z^i_\text{L}$ represent deep representations obtained by models $f^1_{\theta}(\cdot)$ and $f^2_{\phi}(\cdot)$ for $\bm X_{L}$ and $\bm X_{UL}$, respectively. Incorporating the InfoNCE loss aids in generating complex pixel representations by leveraging ample unlabeled data, thereby enhancing the resilience and label efficiency of segmentation models.

\section{Experiments and Results}

\subsection{Datasets}
\textbf{The FUGC dataset:} We utilize one transvaginal dataset in this study. It contains images that include anatomical structures of the cervix, namely the anterior and posterior lips. They are captured using a curved transducer with a frequency range of 2 to 10 MHz, specifically a vaginal probe utilized for cervical ultrasound screening in second-trimester patients. Operators are directed to refrain from applying post-processing techniques or introducing artifacts like smoothing, noise, pointers, or calipers whenever feasible. Other image settings, such as gain, frequency, and gain compensation, are adjusted according to individual discretion.
The training set comprises 500 images, the external validation set comprises 90 images, and the testing set comprises 300 images. Among these 500 images, only 50 images are annotated by experienced operators. The resolutions for all images are $336 \times 544$. During model training, 500 images are resized to the size of $224 \times 224$. We use the 50 labeled images to evaluate the segmentation performance of the dual neural networks and save the best model among them.

\textbf{Data Augmentation:}
In this study, data augmentations are implemented on the labeled dataset. These augmentation techniques include rotation within range $(-20^\circ, 20^\circ)$, random brightness contrast, random blur with probability $\mathcal{P}(\cdot)=0.3$, and gaussian noise with probability $\mathcal{P}(\cdot)=0.3$.

\subsection{Implementation Details}
All experiments were implemented in Pytorch and trained on a single RTX 3090 24G GPU. We use a batch size of 8 for training, including two labeled samples and six unlabeled data samples. We adopted a stochastic gradient descent optimizer with a learning rate 0.001, momentum of 0.9, and weight decay of 0.0001. Besides, we set a maximum of 30,000 iterations for training. All neural networks are initialized with random parameters. The experiments used the same hyperparameter for a fair comparison. We also investigated various hyperparameter settings, including a labeled batch size of 1, an unlabeled batch size of 9, and an initial learning rate of 0.01. However, it was our adopted hyperparameter setting that ultimately delivered optimal performance, characterized by fast convergence.

\begin{figure}[htbp]
  \centering
  \includegraphics[width=\linewidth]{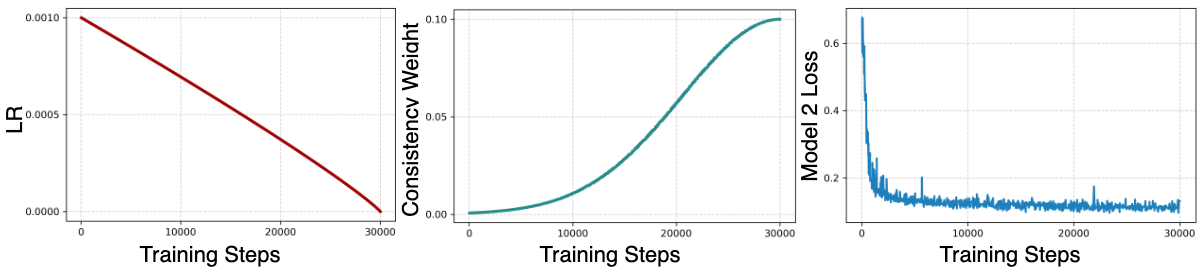}
  \caption{Plots of the hyperparameter settings. From left to right: (a) Learning rate. (b) Consistency weight $\lambda$ in Equation \ref{loss}. (c) Training loss of model $f^2_{\phi}(\cdot)$.
  }  \label{parameters}
\end{figure}

\begin{figure*}[htbp]
  \centering
  \includegraphics[width=\linewidth]{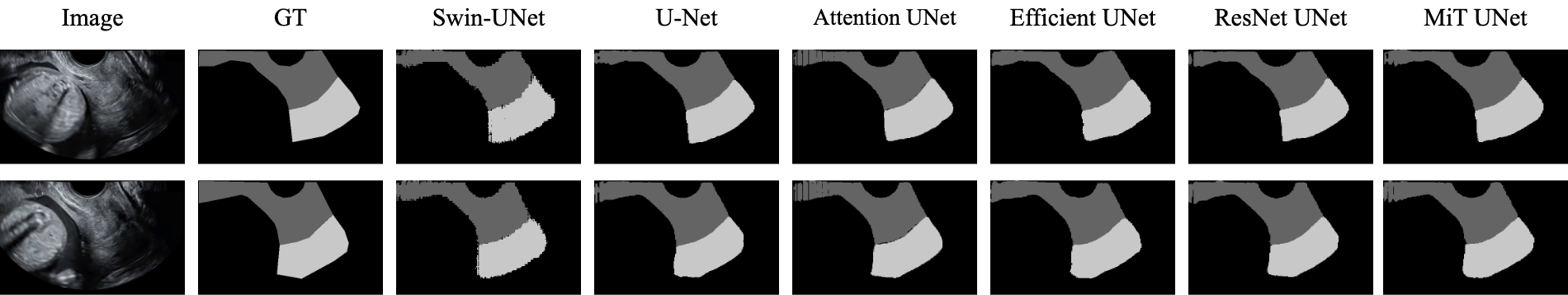}
  \caption{Example ultrasound images from our validation set, ground truth (GT), and corresponding segmentation results of Swin-UNet, U-Net, Attention UNet, Efficient UNet, ResNet UNet, and MiT UNet.}  \label{results_vis}
\end{figure*}

\subsection{Evaluation Metrics}
To measure the performance of our framework, we employ the area metric and the boundary metric: DSC and Hausdorff Distance (HD). The DSC measures the percentage of area overlap between the predicted and ground truth segmentation maps. HD measures the error between the predicted and ground truth segmentation boundaries in pixels. Additionally, we incorporate the execution time to measure the inference speed.

\begin{table}[htbp]
\caption{Segmentation performance of different neural networks on the external validation set.}
\label{multi_networks}
\begin{center}
\setlength{\tabcolsep}{7.3pt}
\begin{tabular}{c|c|c|c|c}
\hline
$\boldsymbol f^1_{\boldsymbol \theta}(\boldsymbol \cdot)$ & $\boldsymbol f^2_{\boldsymbol \phi}(\boldsymbol \cdot)$ & \textbf{DSC} $\uparrow$ & \textbf{HD} $\downarrow$ & \textbf{Time} $\downarrow$ \\
\hline\hline
U-Net& U-Net& 0.76 & 63.99 & 5.65 \\
U-Net& Attention U-Net& 0.78 & 78.68 & 8.89 \\
U-Net& Efficient-Unet& 0.84 & 60.52 & 74.62 \\
U-Net& ResNet-Unet& 0.82 & 69.94 & 7.79 \\
U-Net& MiT-Unet& 0.82 & 62.88 & 7.04 \\
\textbf{U-Net}& \textbf{Swin-Unet} & \textbf{0.86} & \textbf{46.44} & \textbf{16.95} \\
\hline
\end{tabular}
\end{center}
\end{table}

\subsection{Validation}
Fig.~\ref{parameters} illustrates our optimized hyperparameter configuration. The learning rate decays to facilitate the convergence of models $f^1_{\theta}(\cdot)$ and $f^2_{\phi}(\cdot)$ at the completion of training. The consistency weight $\lambda$ guides the models towards learning effectively from the unlabeled training set. The training loss is minimized without additional decays over the course of our maximum epochs.

Table~\ref{multi_networks} illustrates the quantitative results of our SSL framework with several types of UNet, including U-Net \cite{Ronneberger:2015}, Attention U-Net \cite{oktay:2018}, Efficient-Unet \cite{Tan:2021}, ResNet-Unet \cite{He:2016}, MiT-Unet \cite{Xie:2021} and Swin-Unet \cite{Cao:2023} when 50 cases in the training set are labeled data for training. The Swin-UNet outperforms other UNet variants, achieving a DSC of 0.86 and an HD of 46.44. Although the Swin-UNet has a longer inference time than U-Net and Attention U-Net, the value of 16.95 remains within an acceptable range. We computed the average evaluation metrics for models on the validation set to determine the optimal model for future testing. Then we choose the best one for the challenge competition.

In Fig.~\ref{results_vis}, we visualize two randomly selected example ultrasound images from our validation set (50 images). The Swin-Unet model demonstrates highly accurate predictions that closely align with the ground truth. 

\begin{table}[htbp]
\caption{Performance comparison between the baseline method and our proposed framework on the testing dataset. Baseline: A model provided by the challenge organizer.}
\label{test_results}
\begin{center}
\setlength{\tabcolsep}{10pt}
\begin{tabular}{c|c|c|c|c}
\hline
\textbf{Rank} & \textbf{Method} & \textbf{DSC} $\uparrow$ & \textbf{HD} $\downarrow$ & \textbf{Time} $\downarrow$ \\
\hline\hline
1 & N/A & 0.90 & 44.94 & 365.91 \\
2 & N/A & 0.90 & 40.20 & 340.24 \\
3 & N/A & 0.87 & 45.70 & 518.38 \\
\textbf{11} & \textbf{Ours} & \textbf{0.78} & \textbf{66.74} & \textbf{43.51} \\
14 & Baseline & 0.72 & 115.41 & 402.30 \\
\hline
\end{tabular}
\end{center}
\end{table}

\subsection{Comparison with Other Methods}
In Table~\ref{test_results}, we compare our method to the top 3 methods and the baseline method on the testing dataset provided by the challenge organizer. Compared to the baseline method, our SSL framework increases the DSC by 0.06 while reducing the inference time by 82\%. Moreover, the HD significantly decreased from 115.41 to 66.74.

\section{Conclusion}
We conducted a study to explore the implementation of UNet-based architectures within a semi-supervised approach for cervical segmenting on ultrasound images. We introduced an innovative learning strategy that integrates cross-supervision and contrastive learning techniques to optimize the performance of SSL. Through our experimental analyses, we highlighted the efficacy of our SSL framework.
In future studies, we aim to enhance our research by refining our methodologies in restricted supervised learning scenarios while persisting in utilizing the distinctive features offered by UNet variants.



\section{Acknowledgment}
This publication has emanated from research conducted with the financial support of Taighde \'{E}ireann – Research Ireland under Grant number 18/CRT/6183. For the purpose of Open Access, the author has applied a CC BY public copyright licence to any Author Accepted Manuscript version arising from this submission

\bibliographystyle{IEEEbib}
\bibliography{strings,refs}

\end{document}